\documentclass[prl,twocolumn,twoside,showpacs,floatfix]{revtex4}
\usepackage{epsf,amssymb,mathptmx}
\begin{document}
\noindent [Phys. Rev. Lett. {\bf 91}, 058701 (2003)]
\title{Probabilistic prediction in scale-free networks: Diameter changes}
\author{J.-H.\ Kim, K.-I.\ Goh, B.\ Kahng and D.\ Kim \\}
\affiliation{
\mbox{School of Physics and Center for Theoretical Physics,
Seoul National University, Seoul 151-747, Korea}}
\date{\today}
\begin{abstract}
In complex systems, responses to small perturbations are too
diverse to predict how much they would be definitely,
and then such diverse responses can be predicted in a probabilistic way.
Here we study such a problem in scale-free networks,
for example, the diameter changes by the deletion of a single vertex
for various {\it in silico} and real world scale-free networks.
We find that the diameter changes are indeed diverse and their
distribution exhibits an algebraic decay with an exponent
$\zeta$ asymptotically.
Interestingly, the exponent $\zeta$ is~robust as $\zeta \simeq 2.2(1)$
for most scale-free networks, insensitive to the degree
exponents $\gamma$ as long as $2 < \gamma \le 3$. However, there is
another type with $\zeta \simeq 1.7(1)$ and
its examples include the Internet and its related {\it in silico} model.

\end{abstract}
 \maketitle A complex system
consists of many constituents, generating emerging behaviors
through diverse interactions \cite{nature,science}. One of the
powerful ways of examining the intrinsic nature of a complex
system is to observe how such emerging patterns change by small
perturbation applied to the system. In complex systems, such a
change or response is so sensitive to the details of the
perturbation that it is extremely diverse. In such a case, it is
not adequate to {\it predict} how much the change would be
definitely. Recently, Parisi has argued~\cite{parisi} that the
prediction for the responses to small perturbations in complex
systems can be made in a probabilistic way. He showed examples of
protein structures in biological systems and spin glasses in
physical systems. In case of proteins, subject to small external
perturbations such as the change in pH or the substitution of a
single amino acid, they would fold to a completely different 3D
structure but with practically the same free energy.
In case of the disordered magnetic
systems, each spin responds to a slowly varying external field by
changing its orientation, forming a series of bursts, known as
Barkhausen noise~\cite{crackle}. The number of spins burst depends
on the disorder strength of the system, following a power-law
distribution at a critical strength of disorder. The prediction of
the number of spins burst in this case can only be probabilistic.
The stock market is another example of complex systems.
Stock prices are determined as a result of the complicated
interplay between numerous investors, and the price changes were
also found to exhibit a power-law distribution~\cite{price}. All
these examples aptly illustrate how the concept of probabilistic
prediction may apply as a new paradigm in modern science. Other
examples can also be found in as diverse fields as the meteorology
and the geology~\cite{buchanan}.

Recently, there are many works to describe complex systems
in terms of graphs~\cite{rev,porto}, where vertices represent
constituents and edges interactions between constituents.
An interesting feature emerging in such complex networks
is the emergence of a power-law behavior in the degree distribution,
$P_{\rm d}(k) \sim k^{-\gamma}$~\cite{ba},
where the degree $k$ is the number of edges incident upon a
given vertex. Such complex networks
are called scale-free (SF) networks.

In this Letter, we study how SF networks respond to small
perturbations and check if the concept of probabilistic prediction
can be applied. For this purpose, we investigate a simple problem
of diameter change when a single vertex is removed from the
system. Diameter, defined as the average distance between every
pair of vertices in a network, is a simple yet fundamental
quantity of SF networks to characterize the small-world nature,
and can be thought of as a measure reflecting the efficiency of a
network. Our main interest is how much the efficiency of a network
would be affected by the removal of a single vertex. When a vertex
is removed, each pair of remaining vertices whose shortest pathway
had passed through the removed vertex should find detours,
resulting in the rearrangement of shortest pathways over the
network. Thus the diameter change occurs in a collective manner.
From the extensive numerical calculations for a number of SF
network models and real-world examples, we find that the diameter
changes indeed are very diverse and crucially depend on the degree
of the removed vertex. When a vertex with a few number of
connections is removed, the diameter changes little. However, when
a vertex with a large number of connections is removed, the
diameter change is drastic, exhibiting a power-law distribution
with an exponent $\zeta$,
\begin{equation}
P_{\rm c}(\Delta) \sim \Delta^{-\zeta}
\label{zeta}
\end{equation}
for large $\Delta$, where $\Delta$ is the dimensionless relative diameter change
defined as the diameter change caused by the removal of a
certain vertex divided by the original diameter before the removal,
and $P_{\rm c}(\Delta)$ is its distribution.
Moreover the exponent $\zeta$ turns out to be robust
for various SF networks, insensitive to the degree exponent $\gamma$ for $2<\gamma\le 3$.

To be specific, we consider a undirected SF network with finite
number of vertices $N$ and measure the diameter of the network.
Note that we limit our interest to undirected networks only in
this work.
Next we remove a certain vertex $i$ and measure the
diameter $d_i$ of the rest of the network. Measuring a
dimensionless quantity, $\Delta_i=(d_i-d_0)/d_0$ for all $i$,
where $d_0$ is the diameter of the original unperturbed network,
we obtain the distribution of $\Delta$ for the network. Note that
our case is different from the previous study of the robustness of
SF networks~\cite{archilles,lai} where vertices are removed
successively. In our case, on the other hand, only a single vertex
is removed each time. When a certain vertex is removed, the
network may disintegrate into more than one cluster. In such
cases, $d_i$ is calculated only within the largest cluster. The
diameter can be measured via a simple breadth-first search
algorithm. To obtain the distribution of the diameter changes, we
need the computation time of order ${\cal O}(N^3)$.

To look into more details, we consider the static model
\cite{load}.
{It is constructed by connecting $mN$ pairs of
vertices $i$ and $j$ with probability proportional to
$(ij)^{-\alpha}$, where $N$ is the vertex number and $\alpha$ is a
parameter. We use $m=2$. Its degree distribution follows a power
law, $P_d(k) \sim k^{-\gamma}$ with $\gamma=1+1/\alpha$. Thus,
tuning the parameter $\alpha$ in $[0,1)$, we obtain a continuous
spectrum of the exponent $\gamma$ in the range $2<\gamma<\infty$.
}

\begin{figure}[t]
\centerline{\epsfxsize=8.cm \epsfbox{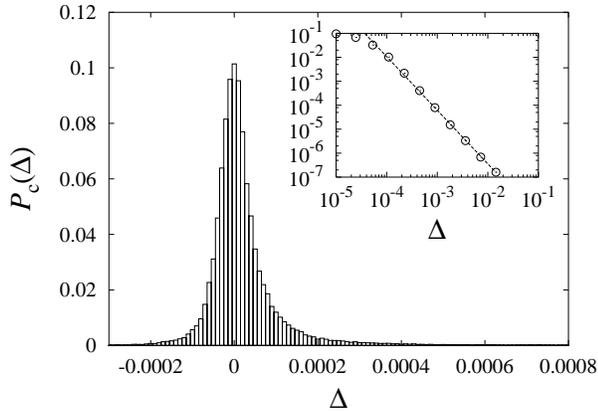}}
\caption{Normalized histogram of the diameter changes for
the static model with $\gamma=3$ and $N=10^4$, averaged over 10 configurations.
Horizontal range
is truncated for clearance, but runs up to $2\times10^{-2}$.
Inset: Plot of $P_{\rm c}(\Delta)$ in log-log scale for $\Delta>0$.
Dashed line is a fit line having a slope $-2.2$.
Data points are logarithmically binned.}
\label{histogram}
\end{figure}

\begin{table*}
\caption{Summary of the results for various SF networks. Tabulated for each network are the system size $N$, the mean degree $\langle k\rangle$, the degree exponent $\gamma$, the diameter change exponent $\zeta$, and the betweenness centrality exponent $\eta$ \cite{class}.
}
\begin{ruledtabular}
\begin{tabular}{rlcccccc}
 & System & $N$ & $\langle k\rangle$ & $\gamma$ & $\zeta$ & $\eta$ & ref.\ \\
\hline
(i) & Static model & $10^4$ & 4 & 2.2--3.0 & 2.2(1) & 2.2(1) & \cite{load}\\
(ii) & Barab\'asi-Albert model & $10^4$ & 4 & 2.2--3.0 & 2.2(1) & 2.2(1) & \cite{ba}\\
(iii) & Copying model & $10^4$ & 4 & 2.2--3.0 & 2.2(1) & 2.2(1) & \cite{kumar}\\
(iv) & Fitness model & $10^4$ & 4 & 2.25 & 2.2(1) & 2.2(1) & \cite{fitness}\\
(v) & Accelerated-growth model & $10^4$ & ${\cal O}(1)$ & 3.0(1) & 2.2(1) & 2.2(1) & \cite{accel}\\
(vi) & Huberman-Adamic model & $10^4$ & ${\cal O}(1)$ & 3.0(1) & 2.2(1) & 2.2(1) & \cite{ha}\\
(vii) & Protein interaction network model & $10^4$ & ${\cal O}(1)$ & -- & 2.2(1) & 2.2(1) & \cite{sole_pin}\\
(viii) & Protein interaction network of the yeast {\it S.~cerevisiae}  & $5662$ & 6.1 & 3.2(2) & 2.3(1) & 2.3(1) & \cite{goh3}\\
(ix) & Internet at the autonomous systems level & $6474$, $12058$ & $\sim$4 & 2.1(1) & 1.7(1) & 2.0(1) & \cite{nlanr}\\
(x) & Adaptation model & $\sim$$6500$ & ${\cal O}(1)$ & 2.1 & 1.7(1) & 2.0(1) & \cite{goh_internet}\\
\end{tabular}
\end{ruledtabular}
\label{table}
\end{table*}

The diameter change $\Delta_i$ by the removal of a certain vertex
$i$ in such SF networks can be positive or negative, and the
histogram of the diameter changes is highly centralized around
$\Delta=0$ (Fig.~1). However, it exhibits a fat tail for $\Delta >
0$ (the inset of Fig.~1). For the static model with $\gamma=3$,
for example, the case of small diameter changes in the range
$|\Delta | < 2 \times 10^{-4}$ occurs with frequency as high as
96\%. Thus the effect of a vertex removal usually is negligible as
a whole, which is manifested by the exponentially bounded
fluctuations of the diameter around its original value. We
estimate the $N$-dependence of such small diameter changes in a
mean-field-type approach. It is known that the diameter $d_0$
depends on the number of vertices as $d_0 \sim \ln N$ for random
graph and $d_0 \sim \ln N / \ln \ln N$ \cite{ultra} for the
Barab\'asi-Albert model \cite{ba} with $\gamma=3$. When a vertex
is removed, the diameter may be reduced as $d \sim \ln (N-1)$ or
$d \sim \ln (N-1)/ \ln \ln (N-1)$, both leading to $\Delta \approx
-1/N \ln N$ for large $N$. Thus when $N=10^4$, $\Delta \sim {\cal
O}(10^{-5})$, which is comparable to numerical values of the
central part in Fig.~1. On the other hand, substantial (about 4\%)
vertices have a serious impact on the system's efficiency and they
indeed contribute to the positive tail of the histogram, showing
the power-law behavior, Eq.~(\ref{zeta}). We find that such large
diameter changes are mainly due to the removal of a vertex with
large degree. { This feature is reminiscent of the percolation
problems on the SF networks~\cite{percol,havlin}. 

\begin{figure}[t]
\centerline{\epsfxsize=8.cm \epsfbox{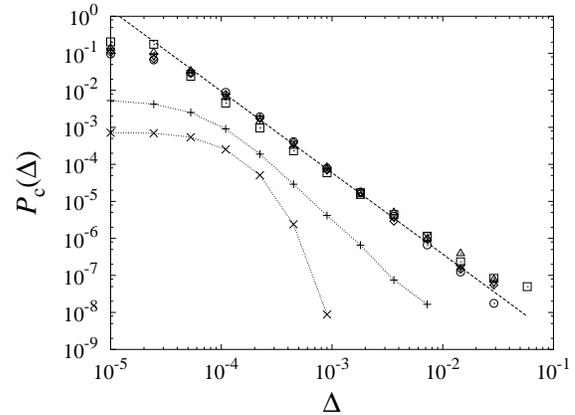}}
\caption{The diameter change distribution $P_{\rm c}(\Delta)$ for
the static model with $\gamma=2.2$ ($\Box$), 2.4 ($\Diamond$),
2.6 ($\bigtriangledown$), 2.8 ($\triangle$), 3.0 ({\Large$\circ$}), and $4.0$ ($+$), and the Erd\"os-R\'enyi model ($\times$).
The data, obtained for $N=10^4$ and averaged over $10$ configurations.
The two data sets $(+, \times)$ are shifted vertically for comparison.
Dashed line having a slope $-2.2$ is drawn for the eye.
Note that the deviations from the straight line
at the fat tail are due to the generic finite-size effects
for the SF networks with $\gamma<3$~\cite{finiteness}.
}
\label{static}
\end{figure}

Let us investigate the power-law behavior for large $\Delta$
in details. The exponent $\zeta$ seems to be robust as
$\zeta \approx 2.2(1)$ as long as $2 < \gamma \le 3$ for the
static model as shown in Fig.~2.
Similar behaviors are found in other model networks (ii)--(vii)
listed in Table 1.
These include the SF networks showing nontrivial degree-degree
correlations \cite{assort}.
For $\gamma>3$, on the other hand, as $\gamma$ increases,
the power-law behavior sets in only for larger values of $\Delta$ and
the exponent $\zeta$ increases with $\gamma$.
Eventually the diameter change distribution
for the Erd\"os-R\'enyi random networks decays exponentially
as shown in Fig.~2.

To see such universal behavior of $\zeta$ in real world,
we consider a couple of real-world networks, the protein
interaction networks (PIN) and the Internet.
For the PIN of the yeast
{\it Saccharomyces cerevisiae}~\cite{goh3},
we also find a power law in the diameter change distribution
with an exponent $\zeta\approx2.3(1)$ (Fig.~3), consistent with the one
obtained for various model networks, including the one
proposed as its own {\it in silico} model (vii) \cite{sole_pin}.
\begin{figure}[b]
\centerline{\epsfxsize=8.cm \epsfbox{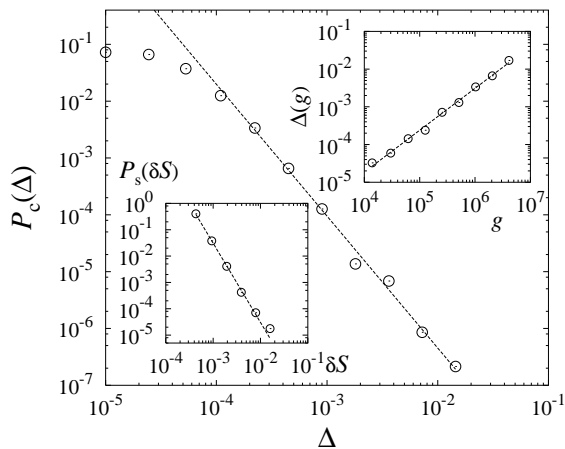}}
\caption{The diameter change distribution $P_{\rm c}(\Delta)$ for the
PIN of the yeast {\it S.~cerevisiae}.
The slope of the fit line (dashed) is $-2.3$, drawn for the eye.
Upper inset: Plot of $\Delta(g)$ {\it vs.} $g$. The slope of the straight
line is $1.1$, drawn for the eye.
Lower inset: The largest-cluster-size change distribution
$P_{\rm s}(\delta S)$. Here $\delta S$ is normalized by $N$.
The slope of the fit line is $-3.0$,
drawn for the eye.
}
\label{pin}
\end{figure}
For the Internet at the autonomous systems level~\cite{nlanr},
the diameter change distribution again
follows a power law, however, with a different exponent
$\zeta \approx 1.7(1)$ (Fig.~4). The smaller exponent $\zeta$ indicates that
the effect of the removal of vertices contributing to
the tail of the distribution is much more severe
than the previous cases with $\zeta\approx2.2$ [(i)--(viii) in Table 1].
To confirm the novel value of $\zeta$ for
the Internet, we perform the same calculations for its {\it in silico}
model, called the adaptation model \cite{goh_internet}, and
indeed obtain $\zeta\approx1.7$ for it, too. The two different behaviors
of the diameter change distribution are rooted from distinct topological
features of shortest pathways of each case, which will be discussed
later.
\begin{figure}[b]
\centerline{\epsfxsize=8.cm \epsfbox{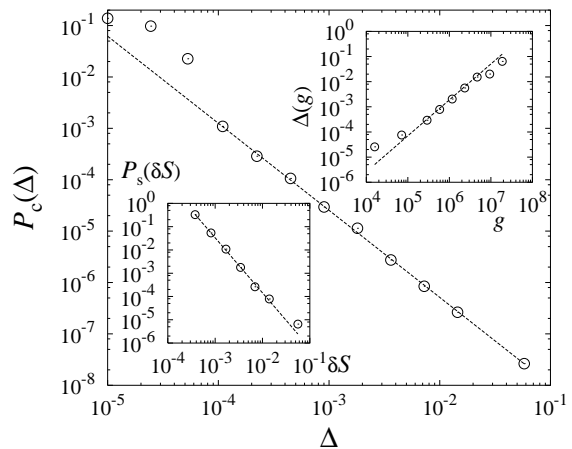}}
\caption{The diameter change distribution $P_{\rm c}(\Delta)$ for the
Internet at the autonomous system level.
The slope of the fit line (dashed) is $-1.7$, drawn for the eye.
Upper inset: Plot of $\Delta(g)$ {\it vs.} $g$. The slope of the straight
line is $1.4$, drawn for the eye.
Lower inset: The largest-cluster-size change distribution
$P_{\rm s}(\delta S)$. Here $\delta S$ is normalized by $N$.
The slope of the fit line is $-2.4$, drawn for the eye.
}
\label{internet}
\end{figure}

Recently, it was proposed that the SF networks with $2<\gamma\le3$
can be classified into two classes~\cite{load,class}, following
the power-law behavior of the betweenness centrality (BC)
distribution~\cite{freeman,newman1}. { The BC $g_k$ of a vertex
$k$ is the accumulated sum of the fraction of shortest pathways
passing through $k$ and its distribution follows a power law,
$P_g(g) \sim g^{-\eta}$ for SF networks.}
The BC exponent $\eta$ turns out to be
robust as either $\eta \simeq 2.2(1)$ (class I) or $\eta\approx
2.0(1)$ (class II) as long as $2 < \gamma \le
3$~\cite{load,class}. Interestingly, the networks (i)--(viii) in
Table 1 having the diameter change exponent $\zeta\approx2.2$
belong to the class I, and the values of $\zeta$ and $\eta$
coincide with each other within our numerical resolutions, while
they are different for the class II. Empirically, the rank of a
vertex in $g$ and that in $\Delta$ are likely to be the same for
vertices with large degrees. If then, the relation $P_g(g) dg \sim
P_c(\Delta) d\Delta$ would hold asymptotically, leading to
\begin{equation}
\Delta(g)\sim g^{(\eta-1)/(\zeta-1)},
\end{equation}
for large $g$. This type of relation also holds between degree
and BC~\cite{load}.
Indeed, the slopes in the double logarithmic scale in the upper
insets of Figs.~3 and 4
are $1.1(1)$ for the PIN and $1.4(1)$ for the Internet, respectively,
consistent with the predictions from the formula, Eq.~(2).
Thus the two classes, the classes I and II, are also categorized by
the diameter change distribution and the distinction between them
can be observed more clearly through it.

Our finding that the diameter change distribution is also
classified into the classes I and II following those for BC
distribution may be rooted from the fact that both quantities,
diameter and BC, depend on universal features of {\it the shortest
pathways} topology between a vertex pair in networks. When the sum
rule \cite{sumrule}, {\it i.e.}, $\sum_k g_k \sim d$ is applied,
one can see immediately that the diameter change distribution is
the same as the total BC change distribution. On the other hand,
the networks belonging to the class II are more sparse and
ramified than those in the class I, so that the Internet is more
fragile by the removal of a single vertex than the PIN. We compare
the distribution of the size change $\delta S$ of the largest
cluster for the PIN and the Internet by a single vertex removal.
As shown in the lower insets of Figs.~3 and 4, the giant cluster
in the Internet becomes much smaller than in the PIN. Thus the
number of vertex pairs connected
after the removal becomes much smaller in the Internet
than in the PIN. Consequently, the difference of the exponent
$\zeta$ between the two classes appears much larger than that of
the exponent $\eta$ in the class II. However, it is not clear how
the power-law behavior in $P_c(\Delta)$ arises and what determines
its exponent.

It would be interesting to generalize our study to the case of having more
than one vertex removed. For simplicity, we consider the case of
two-vertex removal, in particular, one is the hub and the other
is an arbitrary vertex. Interestingly we find that the diameter change
distribution also exhibits a fat-tail behavior with the same exponent
$\zeta$. We cannot check if the fat-tail behavior still holds
for more general cases due to huge amount of computing time.
Meanwhile, it has been studied that SF network
is robust against random failures. To show this, the diameter
change due to removal of a finite fraction of vertices was measured
by taking average over a few samples, not over the whole ensemble.
If the diameter change distribution after those removals still possesses
a power law distribution with $\zeta \le 3$, as is the case here,
then we could say that the average diameter change cannot reflect
its intrinsic nature because its variance diverges. Thus
the diameter changes after random failures should also
be described in a probabilistic way.

In summary, we have studied the diverse behavior in response to
a small perturbation, a deletion of a single vertex in SF networks.
The diameter change $\Delta$ by a removal of a vertex is very diverse,
exhibiting a power-law distribution with an exponent $\zeta$ for
large $\Delta$.
Moreover, the diameter change exponent $\zeta$ is robust as
$\zeta\simeq 2.2$ for most SF networks with $2<\gamma\le3$,
or $\zeta\simeq1.7$ for the Internet as an exception.

\begin{acknowledgments}
This work is supported by the KOSEF Grant No. R14-2002-059-01000-0
in the ABRL program.
\end{acknowledgments}

\end{document}